\documentclass[aps,twocolumn,showpacs,amsmath,amssymb,superscriptaddress]{revtex4-1}

\usepackage{graphicx}
\usepackage{color}
\usepackage{subfigure}
\usepackage{dcolumn}
\usepackage{bm}
\usepackage{epsf}
\usepackage{amsmath,amstext,amssymb}
\usepackage{enumitem}
\usepackage{hyperref}
\usepackage[ansinew]{inputenc}
\usepackage{braket}
\usepackage{multirow}
\PassOptionsToPackage{numbers,sort&compress}{natbib}

\makeatletter
\g@addto@macro\normalsize{%
  \setlength\abovedisplayskip{8pt}
  \setlength\belowdisplayskip{8pt}
  \setlength\abovedisplayshortskip{8pt}
  \setlength\belowdisplayshortskip{8pt}
}
\makeatother

\newcommand{\be}{\begin{equation}}
\newcommand{\ee}{\end{equation}}
\newcommand{\bea}{\begin{eqnarray}}
\newcommand{\eea}{\end{eqnarray}}

\begin{document}

\title{A long-lived Zeeman trapped-ion qubit}

\author{T.~Ruster}
\author{C.~T.~Schmiegelow}\thanks{Present address: LIAF -  Laboratorio de Iones y Atomos Frios, Departamento de Fisica \& Instituto de Fisica de Buenos Aires, 1428 Buenos Aires, Argentina}
\author{H.~Kaufmann}
\author{C.~Warschburger}
\author{F.~Schmidt-Kaler}
\author{U.~G.~Poschinger}\email{poschin@uni-mainz.de}

\affiliation{Institut f\"ur Physik, Universit\"at Mainz, Staudingerweg 7, 55128 Mainz, Germany}

\begin{abstract}
We demonstrate a coherence time of 2.1(1)~s for electron spin superposition states of a single trapped $^{40}$Ca$^+$ ion. The coherence time, measured with a spin-echo experiment, corresponds to residual rms magnetic field fluctuations $\leq$~2.7$\times$10$^{-12}$~T. The suppression of decoherence induced by fluctuating magnetic fields is achieved by combining a two-layer $\mu$-metal shield, which reduces external magnetic noise by 20 to 30~dB for frequencies of 50~Hz to 100~kHz, with Sm$_2$Co$_{17}$ permanent magnets for generating a quantizing magnetic field of 0.37~mT.  Our results extend the coherence time of the simple-to-operate spin qubit to ultralong coherence times which so far have been observed only for magnetic insensitive transitions in atomic qubits with hyperfine structure.
\end{abstract}


\maketitle

\section{Introduction}

Quantum technology based on trapped-ion quantum bits has seen steady progress in the past decades. Quantum algorithms and simulations of increasing complexity have been recently demonstrated \cite{MONZ2016, DEBNATH2016, JURCEVIC2014}. A crucial prerequisite is a sufficiently slow decay of qubit coherence. Widely employed optical or spin qubits based e.g. on $^{40}$Ca$^+$ or $^{88}$Sr$^+$ feature a qubit frequency which is linearly dependent on the ambient magnetic field, such that field fluctuations lead to dephasing and thus the stability of the magnetic field becomes crucial. Moreover, it has been shown that entangled states may exhibit an increased sensitivity to magnetic field fluctuations, scaling with the squared number constituent qubits in the worst case \cite{MONZ2011}. Using various technical measures such as $\mu$-metal shielding, active magnetic field stabilization, synchronization to the ac mains and improved current drivers for supply of coils for generation of a quantizing magnetic field leads to typical coherence times of 10-40~ms.\\
By contrast, ion species with hyperfine structure such as $^9$Be$^+$ \cite{BOLLINGER1985}, $^{43}$Ca$^+$ \cite{BENHELM2008,HARTY2014} or $^{171}$Yb$^+$ \cite{OLMSCHENK2007,TIMONEY2011}, allow for encoding quantum information in magnetic field insensitive transitions, which feature a vanishing first-order Zeeman shift. Utilizing these species leads to additional challenges: A more complex level structure can lead to increased sophistication of qubit operation. Furthermore, such transitions require a specific {\it magic} magnetic field, which can restrict the range of possible applications. Some species require large magnetic fields, yielding Zeeman splittings larger than the natural linewidths of cycling or repump transitions. This in turn leads to increased complexity of e.g. Doppler cooling and qubit state preparation and readout. Moreover, some hyperfine species require laser fields at wavelengths in the UV range which are less convenient to generate and manipulate. While all these challenges have been successfully addressed, see e.g. \cite{HARTY2014}, operating a hyperfine qubits still leads to a resource overhead and increased complexity.\\

\begin{figure}[h!t!p]\begin{center}
\includegraphics[width=0.48\textwidth]{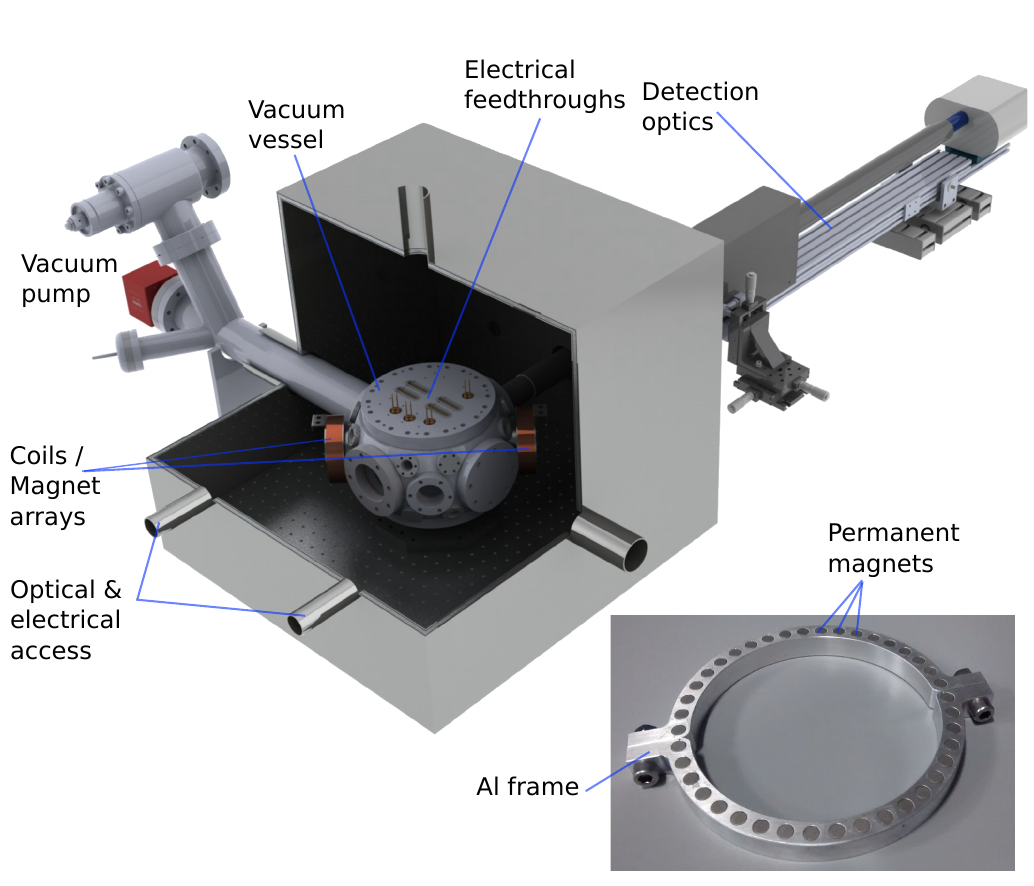}
\caption{Ion trap setup with magnetic shielding enclosure, shown with one top parts removed. The figure does not show the laser focusing optics inside the enclosure. Electrical signals and optical single-mode fibers are supplied via the holes with cylindrical sleeves. The inset shows one of the Al frames bearing the permanent magnets, which are coaxially placed near the coils.
}
\label{fig:aufbau}
\end{center}\end{figure}

A second option to avoid magnetic-field induced decoherence is to employ qubits which are encoded in decoherence-free subspaces of several physical qubits, e.g. Bell states of the type $\ket{01}\pm\ket{10}$ for two constituent qubits. While persisting coherence at wait times of up to 20~s has been demonstrated \cite{KIELPINSKI2001,ROOS2004,HAEFFNER2005}, the number of required qubit ions is increased, as well as the complexity of computational gates \cite{MONZ2009,IVANOV2010}. Another technique for the suppression of slow qubit frequency drifts is dynamical decoupling \cite{BIERCUK2009,BARGILL2013}, which comes at the cost of increased control overhead, particularly for scalable architectures. \\

\begin{figure}[h!t!p]\begin{center}
\includegraphics[width=0.48\textwidth]{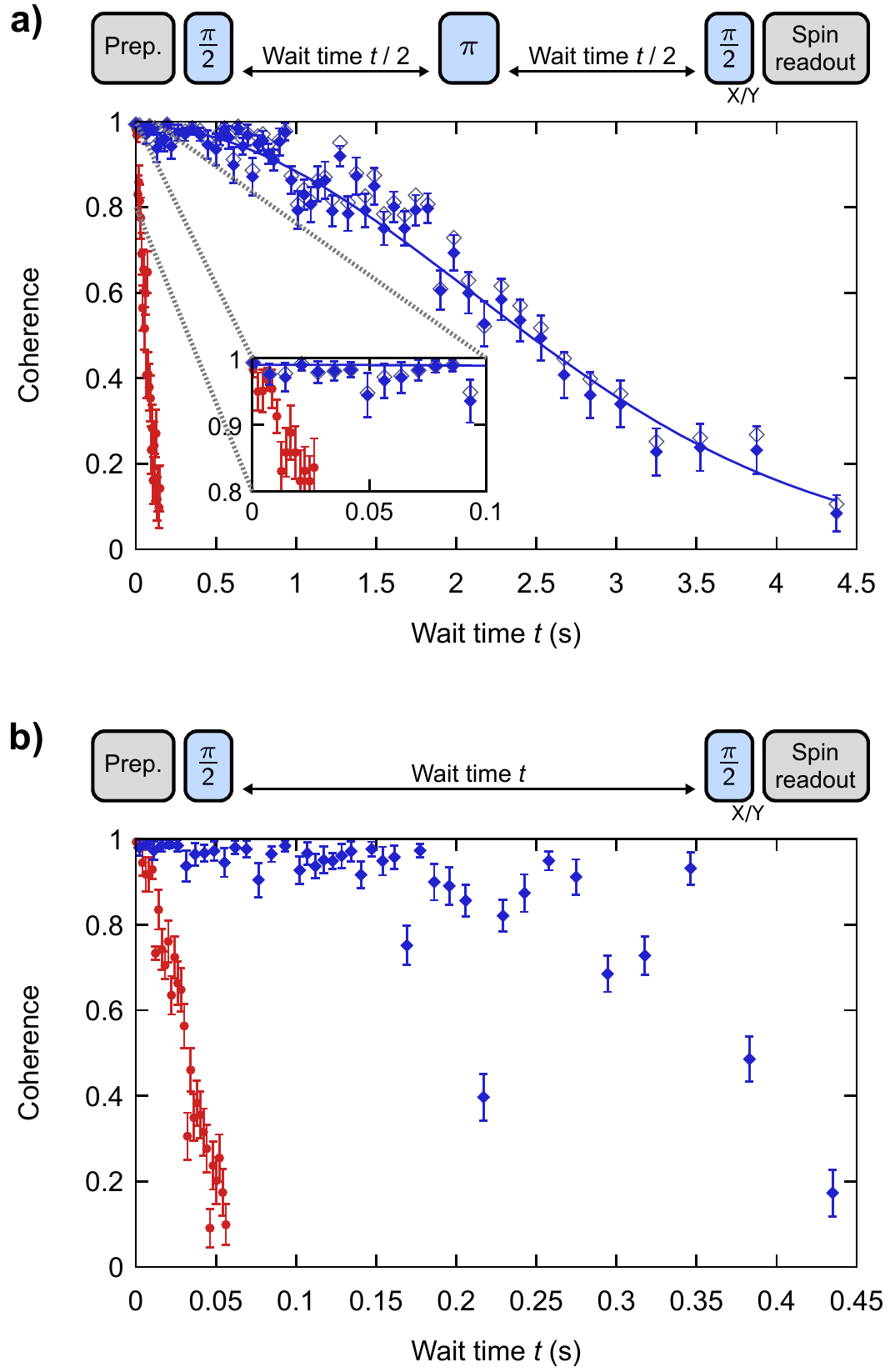}
\caption{Measured qubit coherences versus wait time. \textbf{a)} shows data from spin-echo measurements, while \textbf{b)} shows data from Ramsey measurements. For each panel, the data shown in blue was measured with permanent magnets, whereas the data shown in red pertains to magnetic field generation via coils. In a), the open squares correspond to the permanent magnet data with applied readout error compensation. A Gaussian fit to the permanent magnet spin-echo data with readout compensation reveals a $1/\sqrt{e}$ time of 2.1(1)~s. For each data point, the qubit was measured along $X$ and $Y$ direction 300 times each, and the coherence was inferred via maximum likelihood estimation (see text). The respective measurement sequences are depicted on top of each panel.
}
\label{fig:results}
\end{center}\end{figure}

\section{Experimental setup}

In this work, we demonstrate long qubit coherence times by suppression of ambient magnetic field fluctuations. We trap $^{40}$Ca$^+$ ions in a microstructured segmented Paul trap \cite{SCHULZ2008} and encode quantum information in the Zeeman-split sublevels of the groundstate $\ket{4^2S_{1/2},m_J=+1/2}\equiv\ket{\uparrow}$ and $\ket{4^2S_{1/2},m_J=-1/2}\equiv\ket{\downarrow}$. The qubit is manipulated via stimulated Raman transitions and read out via electron shelving and subsequent detection of state-dependent fluorescence \cite{POSCHINGER2009}. The trap setup, comprised of vacuum vessel, laser focusing optics, trap rf drive resonator and non-evaporable getter pump, is contained within an enclosure of outer dimensions 510~mm (height) $\times$ 625~mm $\times$ 625~mm, see Fig.~\ref{fig:aufbau}. A similar design is reported in \cite{BLAKESTAD2011}. The enclosure \footnote{Sekels GmbH, Ober-M{\"o}rlen, Germany} consists of two layers of $\mu$-metal (an alloy of 80\% Ni and 20\% Fe with a permeability of about 80000), each with 2~mm thickness and separated by a 6~mm Al layer. The enclosure consists of one bottom part and two removable top part, where overlapping $\mu$-metal lids ensure that the shielding efficiency is not compromised by the gaps between the parts. With a pickup coil inside the enclosure, and sending coils outside, we measured attenuation factors in the range between 20-30~dB for signal frequencies between 50~Hz and 100~kHz. All required laser beams are supplied to the trap setup via single-mode fibers. The focusing optics for each beam is remote-adjustable via piezo-controlled mirror holders. All single mode fibers and electrical signals, as well as the free-space photon collection optics, are fed into the enclosure through circular holes with diameters of up to 100~mm. The holes feature cylindrical $\mu$-metal sleeves of 100~mm length outside the enclosure in order to retain shielding efficiency.
The quantizing magnetic field is generated by 80 individual Sm$_2$Co$_{17}$ cylindrical magnets with 6~mm diameter and 4~mm length \footnote{IBS Magnet, Berlin, Germany}, see Fig.~\ref{fig:aufbau}. This material has a remanence of $>$~1~T with a temperature dependence of about -0.03~\%/K - the lowest value known to us for common materials for permanent magnets. The magnets are glued at constant spacing into two round Al frames with an inner diameter of 108 mm and an outer diameter of 128 mm. The two frames are arranged in a quasi Helmholtz geometry, each at a distance of 285~mm from the ion location. This generates a magnetic field of about 0.37~mT at the ion position, leading to a qubit frequency, i.e. Zeeman splitting between $\ket{\uparrow}$ and $\ket{\downarrow}$ of about $2\pi\times$10.5~MHz. This is sufficient to avoid spectral crowding on the S$_{1/2}\leftrightarrow$D$_{5/2}$ quadrupole transition, which is utilized for qubit readout. On the other hand it is smaller than the natural linewidth of the S$_{1/2}\leftrightarrow$P$_{1/2}$ cycling transition, which facilitates Doppler cooling and detection. Eight additional magnets are placed to compensate for the magnetic field gradient at the ion position. This is done by characterizing the magnetic field homogeneity by measuring the Zeeman splitting at different locations along the trap axis \cite{WALTHER2011}, and results in a spatial variation of the Zeeman splitting of less than $2\pi\times$~1~kHz in a position range of $\pm$~1~mm around the trapping location. The closest distance of a magnet to a $\mu$-metal wall is about 20~cm, such that no saturation of the $\mu$-metal occurs, which would compromise the shielding. A further advantage of the permanent magnets  with respect to current-driven coils is the reduction of the heat-load within the enclosure, which was about 15~W in our case. This improves the temperature stability - and therefore the stability of the laser focusing optics - inside the enclosure. \\

\section{Measurement results}
\subsection{Measurement method}

We characterize the qubit coherence via Ramsey-type measurements. Rather than recording complete Ramsey fringes, we measure expectation values of $X$ and $Y$ operators of the electron spin, i.e. we probe the Ramsey signal for two analysis phases at $\pi/2$ difference. While this is sufficient to infer the coherence and phase of a superposition state, it keeps the measurement effort at a minimum, which is particularly relevant for long wait times. Coherent rotations are driven by a pair of co-propagating laser beams near 397~nm, detuned by $2\pi\times$~250~GHz from the $S_{1/2}\leftrightarrow P_{1/2}$ transition. Each beam is individually modulated and switched by a single-pass acousto-optical modulator. The beams are superimposed and jointly supplied to the trap via an optical single-mode fiber. Each measurement sequence starts with a coherent $\pi/2$-rotation on a single Doppler cooled ion initialized in $\ket{\uparrow}$, which results in the superposition state $\tfrac{1}{\sqrt{2}}\left(\ket{\uparrow}+i\ket{\downarrow}\right)$. During a wait time $t$, the superposition accumulates a phase, yielding the state $\tfrac{1}{\sqrt{2}}\left(\ket{\uparrow}+ie^{i\phi}\ket{\downarrow}\right)$. A concluding $\pi/2$ qubit rotation with (without) $\pi/2$ phase w.r.t the first pulse corresponds to qubit readout in the $X$ ($Y$) basis upon projective measurement along $Z$. The estimates of the expectation values obtained from these measurements jointly allow to extract the coherence. For the results presented here, each operator was measured 300 times for a given wait time. The coherence $\vert\rho_{12}\vert$, i.e. the modulus of the off-diagonal element of the density operator, is extracted from the data via maximum-likelihood estimation. For this, we take into account the binomial statistics which govern the projective readout, and we assume balanced population. The latter assumption is justified because the spin qubit exhibits a virtually infinite $T_1$ time, and extrinsic depolarizing processes such as photon scattering do not play a significant role.\\
Note that the maximum likelihood estimation of the superposition coherence is subjected to a bias for the cases with near-unity coherence and superposition phases that are not an integer multiple of $\pi/2$. This leads to unfavorable readout probabilities $0 \ll p_{\ket{\uparrow}} \ll 1$ for \textit{both} $X$ and $Y$ measurements, such that the shot noise increases. Thus, for these cases the maximum-likelihood estimation of $C(\tau)$ yields reduced values with increased error bars.

\begin{figure}[h!tp]\begin{center}
\includegraphics[width=0.5\textwidth]{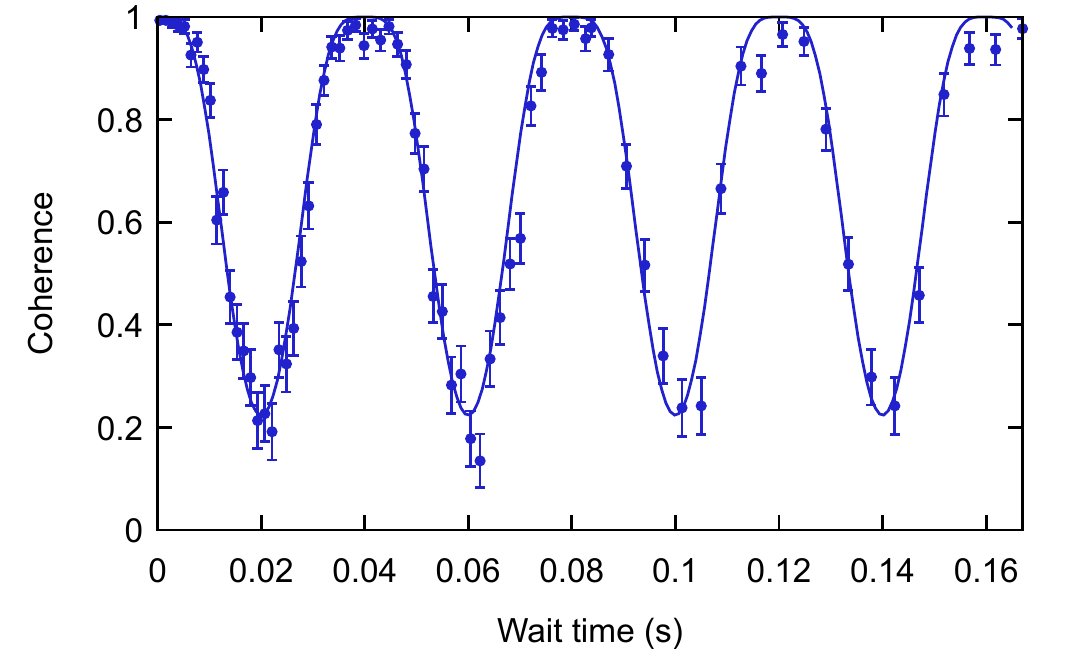}
\caption{Measured qubit coherence versus wait time without triggering to the ac line. The data shows results from spin-echo measurements, and the solid line is a fit to Eq. \ref{eq:contrastwolinetrigger}. The periodic contrast loss is caused by increased phase accumulation from magnetic fields oscillating at the ac line frequency, enhanced by the refocusing pulse at half the line period. The inferred qubit frequency modulation depth is  $\Delta_{ac}=2\pi\times$25.0(5)~Hz. The data was taken with the same method and parameters as in Fig. \ref{fig:results} (see text).
}
\label{fig:woLT}
\end{center}\end{figure}

\subsection{Permanent magnet results}
For each of these measurements, we trigger the sequence on the 50~Hz ac-line, and we perform either a Ramsey measurement, or a spin-echo measurement with an additional refocusing $\pi$-pulse at half the wait time. The latter allows for suppression of fluctuations on a timescale slower than the respective wait time. During the wait time, the ion is shuttled \cite{WALTHER2012} away from the laser interaction region in order to avoid residual scattering of photons on the cooling transition due to imperfect switch-off of the laser near 397~nm driving the S$_{1/2}\leftrightarrow$P$_{1/2}$ cycling transition. The resulting coherences for the different measurement types are shown in Fig. \ref{fig:results}. For the case with refocusing, the contrast exhibits a Gaussian decay $C(\tau)=\exp{(-\tau^2/2\tau_d^2)}$ until wait times $\tau$ beyond 3~s, at a dephasing time constant of $\tau_d$=2.1(1)~s.  The coherence decay exhibits a Gaussian shape, thus, according to Ref. \cite{MONZ2011PHD}, the correlation time of the fluctuations is larger than the maximum measured phase accumulation time. In this limit, the decay time constant corresponds to an rms amplitude of the magnetic field fluctuations of about $\sqrt{\langle \Delta B^2\rangle}=\hbar/(2\mu_B\tau_d)=$2.7~pT. The data pertaining to the cases without refocusing pulse exhibits a more rapid decay $\tau_d^*$=300(50)~ms due to slow fluctuations, presumably caused by drift of the residual penetration of the drifting ambient magnetic field. \\

\subsection{Coil results}
We compare these results to the previous situation where the quantizing magnetic field has been generated by driving currents of 3360~mA through coils with an inner diameter of 75~mm and 170 windings, placed around viewports of the vacuum window and spaced by about 310~mm from the ion location. As the current source, we have employed a specially designed unit \footnote{Design from the Blatt group, University of Innsbruck}, supplied by an \textit{Agilent E3633A} power supply in voltage control mode. Even with the magnetic shielding enclosure closed, we found only a small increase of coherence times to about 30~ms, as compared to 10~ms with the both top segments of the enclosure removed. The corresponding measurement data is shown in Fig. \ref{fig:results}. We conclude that in typical laboratory environment, the impact of ambient magnetic field noise is on the same order of magnitude as the noise generated by low-noise current supplies, such that a substantial enhancement of coherence times requires \textit{both} magnetic shielding and permanent magnets.

\subsection{Residual ac-line feedthrough}
Spin-echo measurements with one or more refocusing pulses allow for measurement of spectral components of magnetic field noise \cite{KOTLER2011}. For the setting with permanent magnets and closed shielding enclosure, we quantify the residual field fluctuations at the 50~Hz ac-line frequency by performing spin-echo measurements without triggering to the ac-line. As can be seen in Fig. \ref{fig:woLT}, periodic loss and revival of coherence is observed at a period corresponding to the ac-line frequency. This occurs because refocusing at half the line period leads to the adverse effect of increase rather the cancellation of the accumulated phase. Assuming sinusoidal modulation of the qubit frequency at a line frequency of $\omega_{ac}=2\pi\times$50~Hz, at a frequency deviation $\Delta_{ac}$ and a random phase at sequence start $\chi_{ac}$, the qubit accumulates a phase
\begin{equation}
\phi_{ac}(\tau,\chi_{ac})=\left(\int_{0}^{\tau/2}dt-\int_{\tau/2}^{\tau}dt\right)\Delta_{ac}\sin(\omega_{ac}t+\chi_{ac}).
\end{equation}
Here, we take into account the inversion of the phase accumulation rate after the refocusing $\pi$-pulse. We assume a uniform distribution of the phase $\chi_{ac}$ of the sequence start w.r.t. the ac-line, which is justified if the sequence duration is not commensurate with the line period. The resulting coherence is then given by averaging over $\phi_{ac}$ with respect to $\chi_{ac}$:
\begin{eqnarray}
C(\tau)&=&\frac{1}{2\pi}\int_0^{2\pi}d\chi_{ac} \exp(i\phi_{ac}(\tau,\chi_{ac}))\\
&=&_0F_1\left(1,-\left(\frac{\pi\Delta_{ac}^2\sin^2(\omega_{ac}\tau/4)}{\omega_{ac}^2}\right)^2\right),
\label{eq:contrastwolinetrigger}
\end{eqnarray}
where $_0F_1\left(a,z\right)$ is the regularized confluent hypergeometric function. Fitting the measured coherence versus wait time to Eq. \ref{eq:contrastwolinetrigger} reveals a residual ac-line induced frequency deviation of $\Delta_{ac}=2\pi\times$~25.0(5)~Hz. For the situation with the magnetic field generated by coils, the frequency deviation was about $\Delta_{ac}=2\pi\times$~300~Hz, while the permanent magnets without shielding enclosure yield $\Delta_{ac}=2\pi\times$~1.5~kHz. Note that for these cases, the frequency deviations were measured by performing Ramsey measurement with short wait times $<$2~ms for fixed phases $\chi_{ac}$ w.r.t. the ac-line, which provides a direct measurement of the variation of the qubit frequency during an ac-line cycle. For comparison, the coherence times and ac-line modulation depths for various settings are summarized in Table \ref{tab:summary}.

\begin{table*}
\centering
    \begin{tabular}{ | c | c | c | c | c | c |}
    \hline
    Field generation & Shield &  Trigger on ac-line  & $\tau_d^*$ time (ms) & $\tau_d$ time (ms) & $\Delta_{ac}/2\pi$ (Hz)\\ \hline\hline
    \multirow{4}{*}{Coils} & \multirow{2}{*}{open} & no  & 0.30(5) & 2.0(2) & \multirow{2}{*}{2500(200)} \\ \cline{3-5}
    &  & yes & 8(1) & 11(2) &  \\ \cline{2-6}
    & \multirow{2}{*}{closed} & no & 1.0(1) & 3.0(2) & \multirow{2}{*}{300(50)} \\\cline{3-5}
    &  & yes  & 28(3) & 45(3) & \\ \hline
    \multirow{4}{*}{Magnets} & \multirow{2}{*}{open} & no & 0.35(5) & 2.0(2) & \multirow{2}{*}{1400(100)} \\ \cline{3-5}
    &  & yes & 17(3) & $>$30 & \\ \cline{2-6}
    & \multirow{2}{*}{closed} & no  & 20(10) & $>$100 & \multirow{2}{*}{25.0(5)} \\ \cline{3-5}
    & & yes & 300(50) & 2100(100) & \\ \hline
    \hline
    \end{tabular}
		\caption{Summary of coherence times and ac-line induced frequency deviations for different experimental settings. The dephasing times $\tau_d^*$ are obtained from Ramsey measurements, whereas the dephasing times $\tau_d$ result from spin-echo measurements. Both $\tau_d^*$ and $\tau_d$ times are reported as $1/\sqrt{e}$ times corresponding to Gaussian decay of the coherence. Note that not for all cases the decay exhibits a Gaussian behaviour, in theses cases the reported values correspond to the time at which the contrast drops below $1/\sqrt{e}$. An exception is the case with permanent magnets, closed shielding enclosure and without ac-line trigger (second-last line, see Fig. \ref{fig:woLT}), where the coherence is periodically reviving even beyond 100~ms.}
		\label{tab:summary}
\end{table*}

\begin{figure}[h!tp]\begin{center}
\includegraphics[width=0.5\textwidth]{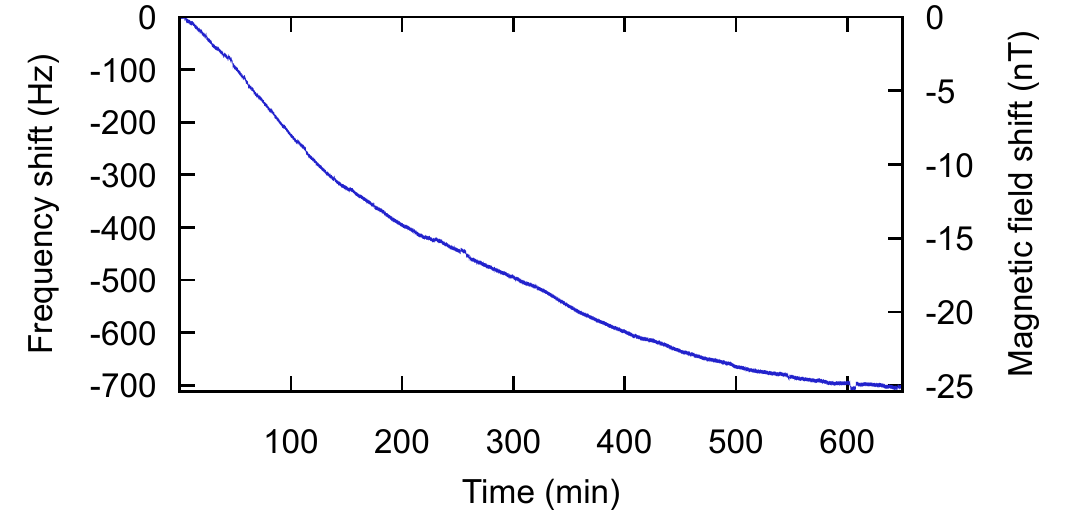}
\caption{Qubit frequency drift: The shift of the qubit frequency for the case with permanent magnets and closed shielding enclosure is tracked over a time of more than 10 hours by a Ramsey measurement at 15~ms wait time. The measurement was started in the evening, and overnight settling can be clearly recognized. The measurement is performed similarly to the measurements presented in Fig. \ref{fig:results} (see text).}
\label{fig:drifts}
\end{center}\end{figure}

\subsection{Long-term stability}
Furthermore, we characterize the long-term drift of the qubit frequency by performing an ac-line triggered Ramsey measurement with $\tau$=15~ms. The qubit frequency is inferred from the extracted phase, and the results are shown in Fig. \ref{fig:drifts}. We observe a maximum drift rate of about $2\pi\times$~0.05~Hz/s, and a total frequency change of about $2\pi\times$~700~Hz over 10 hours. The settling behavior suggests thermal drifts to be responsible for the qubit frequency drift. With the dependence of the remanence of Sm$_2$Co$_{17}$ of -0.03\%/K mentioned above, this would correspond to a temperature change of about -0.2~K within 10 hours.\\

\section{Decoherence sources}

The following mechanisms can cause the remaining decoherence:
\begin{itemize}
\item
Limited shielding efficiency of the $\mu$-metal enclosure: For the case with permanent magnets and ac-line triggering, we observe that ratio of the rms magnetic field fluctuations for the closed enclosure and the opened enclosure lies in the range of about -30~dB. This is consistent with the shielding specification of the of the enclosure, thus residual penetration of fluctuating ambient magnetic fields is most likely the dominating source of decoherence.
\item
Fluctuations of the 50~Hz ac-mains frequency: We observe shifts of several milliseconds of the ac-mains phase at a delay of 2~s w.r.t. an initial trigger flank during an observation time of about 10~min. This is consistent with 0.05~Hz frequency fluctuations of the ac-mains frequency. Together with the peak-to-peak phase modulation depth of about 1~$\pi$ for a spin echo sequence, this leads to significant dephasing within the maximum measured wait time of 4~s. 
\item
Thermal drifts of the permanent magnets which affect their magnetization: While the long-term drift of the quantizing magnetic field has been characterized, see Fig. \ref{fig:drifts}, noise from the magnets on short timescales has not been quantitatively characterized, therefore this could possibly contribute to the observed decoherence.
\item
Drifts of the trapped ion position in conjunction with the magnetic field inhomogeneity: By performing Ramsey measurements in combination with ion shuttling \cite{WALTHER2011}, we observed a gradient of the Zeeman splitting of about $2\pi\times$~8$\cdot$10$^6$~Hz/m. Thus, positions drifts from uncontrolled charging of the trap or its thermal expansion might also contribute.
\item
Drifts of the relative optical phase between the two beams driving the qubit via the stimulated Raman transition: For a similar setup, we have observed phase drifts of the relative optical phase of about 3~$\pi$ within 10 minutes \cite{SCHMIEGELOW2016}. However, as the interferometer area in the present fiber-coupled setup is much smaller and the phase drifts within 4~s are negligible, this mechanism can be excluded.
\item
Deterioration of qubit readout fidelity: We determined a loss of readout fidelity of about 20\% at a wait time of 4~s, induced by heating of the ion, thus this is excluded to be a limiting mechanism. Coherences corrected for readout efficiency deterioration are also shown in Fig. \ref{fig:results}.
\end{itemize}

\section{Conclusion}

In conclusion, we have realized a low cost environment for AMO-physics experiments at variable, intermediate magnetic fields with short term fluctuations of $\sqrt{\langle\Delta B^2\rangle} \leq$~2.7~10$^{-12}$~T. We compare our setup to other experimental environments with low magnetic field noise: For high fields of a few T strength, specially arranged superconducting solenoids \cite{GABRIELSE1988} yield sub-nT rms magnetic field noise in the frequency range of 10 - 200~Hz \cite{BRITTON2015}, while for fully shielded large facilities operating at null field, rms field fluctuations in the femto-Tesla range are attained \cite{PTBSHIELDING}. \\
We have demonstrated to our knowledge unprecedented coherence times of first-order magnetic-field sensitive atomic qubits. The coherence times become competitive with magnetic-field insensitive atomic qubits, or systems such as nitrogen-vacancy centers \cite{BARGILL2013}, or ensembles of Cs atoms in a $^4$He matrix \cite{KANORSKY1996}. In typical laboratory environments, residual current fluctuations in field coils seem to have a similar impact as ambient magnetic field fluctuations, thus long coherence times may be reached using permanent magnets \textit{in conjunction} with proper shielding. 
 
Our results might have influence on the choice of the employed ion species and experimental setups for future trapped-ion experiments. Also, for ion or neutral atom species with clock states, the detrimental impact of higher-order Zeeman shifts on the operation of atomic frequency standards or precision measurements could be mitigated. For high-fidelity quantum information purposes, our $1/\sqrt{e}$ dephasing time of 2.1~s enables on the order of $10^5$ gate operations \cite{GAEBLER2016}, paving the way for scalable implementations of e.g. topological error correction \cite{NIGG2014} or quantum algorithms \cite{MONZ2016}.

\textbf{Acknowledgements} The use of permanent magnets was inspired during a visit of CTS and FSK at Tobias Sch\"atz' labs at Universit\"at Freiburg. We acknowledge earlier contributions of Andr\`e Kesser for the characterization of the shielding properties of the $\mu$-metal enclosure. We further acknowledge helpful discussions with Georg Jacob and Sven Sturm. This  research  was funded by the Office of the Director of National Intelligence (ODNI), Intelligence Advanced Research Projects Activity  (IARPA),  through  the  Army  Research  Office grant W911NF-10-1-0284. All statements of fact, opinion or conclusions contained herein are those of the authors and should not be construed as representing the official views or policies of IARPA, the ODNI, or the U.S. Government.

\bibliography{lit}

\end{document}